\documentclass[runningheads]{llncs}
\usepackage[T1]{fontenc}
\usepackage{graphicx}
\begin{document}
\title{Pupillary activity in areas of interest from visual stimuli for neonatal pain assessment\thanks{}}%Supported by CAPES.
\author{Roberto Magalhães Jr \inst{1}\orcidID{0000-0003-0911-8155}\and
Rafael Orsi \inst{1}\orcidID{0000-0003-4719-0131} \and
Marina Barros\inst{2}\orcidID{0000-0001-6989-3474} \and
Ruth Guinsburg \inst{2}\orcidID{0000-0003-1967-9861} \and
Carlos E Thomaz \inst{1}\orcidID{10000-0001-5566-1963}}
\authorrunning{Roberto Magalhães Jr et al.}
\institute{FEI University Center, Brazil \and
Paulista School of Medicine - Federal University of São Paulo, Brazil
}
\maketitle              % typeset the header of the contribution
\begin{abstract}
This paper compares the pupillary activity index to traditional eye-tracking metrics like the fixation count and duration in assessing neonatal pain. It explores the benefits of incorporating pupillary activity measures to improve methods that lead to an understanding of cognitive processing and performance evaluation. The estimation of cognitive load using pupil diameter typically involves measures relative to a baseline. Instead, we conducted an eye-tracking study using the Low/High Index of Pupillary Activity to evaluate healthcare experts and non-experts analyzing the faces with and without pain from a dataset of newborn faces. This data was recorded by the Tobii TX300 eye-tracking system in a closed room with controlled lighting. Our contribution is to introduce the LHIPA calculation considering the areas of interest segments of the pupil diameter signal. The results suggest that the visual attention reflected by the traditional metrics may not correspond directly to the respective cognitive load for both sample groups of participants.

\end{abstract}
\section{Introdution}
% Eye contact and gaze direction can reveal the visual target of an observer \cite{Kleinke1986} and provide insights into thoughts and intentions. Despite their significance, there is no definitive guide to utilizing eye behavior to investigate emotional and cognitive processes that influence human behavior \cite{Skaramagkas2021}. However, advancements in eye-tracking technology have allowed researchers to analyze visual attention by quantifying various characteristics of eye movements, such as changes in gaze patterns and fixation durations \cite{Ryan2019}. This natural measurement of visual responses provides a strong opportunity for the study of human cognition \cite{Phillips2014}. 

Traditional eye-tracking metrics used to study cognitive load involved examining models based on the number and duration of fixations. Recent improvements in eye-tracking technology have allowed researchers to explore the relationship between pupil diameter and cognitive activity \cite{Mahanama2022}. However, there is still debate regarding the usefulness of pupil diameter compared to other measures. Therefore, this paper aims to compare the pupillary activity index versus count and duration of fixations considering them as metrics to evaluate a neonatal pain assessment cognitive task. By exploring the potential benefits of incorporating pupillary activity measures into existing eye-tracking protocols, we seek to foster a discussion on the underlying mechanisms of cognitive processing and contribute to the development of complementary measures for assessing cognitive load and performance in decision-making tasks.

\section{Non-Baseline-Related Pupillometric Measures}

The estimation of cognitive load using pupil diameter typically involves measures relative to a baseline. A challenge with this approach is that the pupil's sensitivity to illumination levels in the visual stimulus can be confounding. Overall, many studies do not provide illumination measurements but still assume that pupil diameter reflects cognitive load, regardless of the nature of the stimulus. Duchowski \cite{Duchowski2020,Duchowski2018} suggests alternative metrics that consider relative changes in pupil size over time, like the Index of Pupillary Activity (IPA) and, more recently, the Low/High Index of Pupillary Activity (LHIPA).

Basically, IPA is a metric that measures changes in pupil size as an indicator of cognitive or mental workload. Complementary, LHIPA uses the ratio of low frequency (LF) to high frequency (HF) pupil oscillation. This ratio reflects changes in the parasympathetic and sympathetic components of the autonomic nervous system. Pupil constriction is associated with parasympathetic excitation and/or sympathetic inhibition, whereas pupillary dilation is linked to sympathetic excitation and/or parasympathetic inhibition. Unlike the IPA, the LHIPA decreases with increasing cognitive load, as a higher cognitive load is expected to result in a higher frequency response. Duchowski \cite{Duchowski2020,Duchowski2018} provides details of the computation of both the LHIPA and IPA through wavelet analysis.
% \subsubsection{The Index of Pupillary Activity (IPA)} is calculated by decomposing the pupil diameter signal using wavelets and removing coefficients below a certain threshold. The remaining coefficients are then counted to identify sharp variations in pupil diameter, indicating changes in cognitive effort. A low count suggests minimal cognitive effort, while a high count suggests significant cognitive effort. Another measure introduced by Duchowski is the Low/High Index of Pupillary Activity (LHIPA), which uses the ratio of low frequency (LF) to high frequency (HF) pupil oscillation. This ratio reflects changes in the parasympathetic and sympathetic components of the autonomic nervous system. Pupil constriction is associated with parasympathetic excitation and/or sympathetic inhibition, while pupillary dilation is linked to sympathetic excitation and/or parasympathetic inhibition. Unlike the IPA, the LHIPA decreases with increased cognitive load, as a higher cognitive load is expected to result in a higher frequency response. Duchowski provides details of the computation of both the IPA and LHIPA through wavelet analysis \cite{Duchowski2018,Duchowski2020}. 

\section{Material and Methods}
\label{experiment}

To evaluate the LHIPA, we extended an eye-tracking study using a dataset of newborn faces \cite{orsi2004}. The objective was to track the eye movements of experts and non-experts while they were evaluating about presence or absence of pain. 

% More information about this experiment is given below.
 
\subsection{Subjects, Apparatus, and Stimuli}

Volunteers (N=73) for the study were divided into 44 experts  (4 pediatricians and 40 neonatologists, 33.48 ± 7.01 years old) and 29 non-experts (39.82 ± 10.39 years old), respectively with and without training or clinical experience in pain assessment. 

The data was recorded by the Tobii TX300 eye-tracking system (300Hz) in a closed room with controlled lighting, positioned outside the participant's visual field. Calibration and data acquisition were performed using an auxiliary computer with Tobii Studio control software and eye-tracking equipment.

The study utilized a database of facial images, authorized with consent from family members or guardians, and the project for constructing and analyzing these images was approved by the Ethics Committee for Research of the Federal University of Sao Paulo (UNIFESP), under the numbers 1299/09 and 3.116.146. All data were collected at the Hospital of Sao Paulo (HSP), a university-affiliated hospital of UNIFESP. 

\subsubsection{The experimental procedure} involved presenting 20 images representing 10 newborns, with each pair consisting of a resting image and an image taken after a painful procedure \cite{orsi2004}. Participants completed familiarization tests before evaluating the facial images presented randomly for 7 seconds each. A central fixation point was used to ensure consistent eye-tracking strategies, disregarding fixations within the first 300ms. Each facial image was rated using a numerical analog scale, and the experiment lasted approximately 5 minutes per volunteer.
\subsubsection{To analyze pupillary response} by Area of Interest (AOI) and participant's groups, we applied the Tobii I-VT filter \cite{olsen} in pupil diameter signal to fill in gaps where valid data is missing. After that, the remaining gaps were filled using linear interpolation \cite{orsi2004}. According to Orsi et. al \cite{orsi2004}, 2s exposure to a facial expression is sufficient to make this assessment. Therefore, we consider only this time period to analyze.

Similarly to Duchowski \cite{Duchowski2018}, we used a Daubechies-21 wavelet resulting in a 42x3=126 ms sampling window recommended to LHIPA for data sampled at 300 Hz. Our contribution is to calculate the LHIPA on segments of the pupil diameter signal. We have segmented this according to each AOI (Right and Left Eye, Region between Eyebrows, Forehead, Mouth, Right and Left Nasolabial Groove, Chin, Right and Left Eyebrow, Nose, Right and Left Cheek). After that, we added the result for each eye-tracking visit in the same AOI, as in the traditional approach for the gaze duration and fixation count metrics. Then, we normalized it by visits, because the volunteers did not look at all AOI in the first 2 seconds.

\section{Results}

In our study,  the dependent variable was LHIPA, and we included two categorical independent variables in an Ordinary Least Squares regression model \cite{Dismuke2006}: subject groups (experts and non-experts) and AOIs, including interaction terms between them. The results showed that the groups' coefficients are significant (p-value < 0.05). The interaction terms between the groups and the AOIs are significant for most levels which means that there is also evidence that the relationship between the dependent variable and the independent variable is different between the groups of participants, depending on the area of interest in question. Lastly, the coefficients for AOI are significant (p-values < 0.05 for all of them), revealing that there is more cognitive activity in some areas than in others. Curiously these areas are not the same as those that had the highest values for the number and time of fixations.

These findings are summarized in Fig.~1, which shows the pupillary activity and visual attention by the normalized LHIPA, gaze duration, and fixation count in each AOI for all participants. LHIPA is inversely proportional to the cognitive load, so we used its complement instead (LHIPA*).

\begin{figure}[t] 
\label{fig:map}\vspace*{4pt}
%\centerline{\includegraphics{fx1}\hspace*{5mm}\includegraphics{fx1}}
\centerline{\includegraphics[width=4.8in]{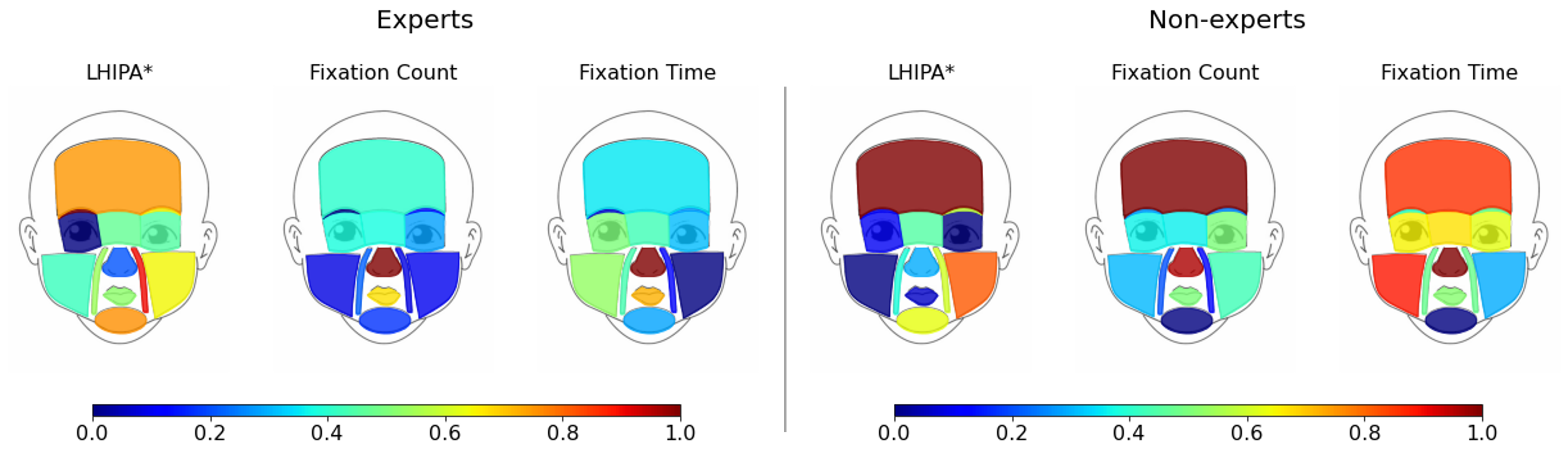}}
\caption{Maps of gaze metrics according to areas of interest and groups of participants.}
\end{figure}

\section{Conclusion}

Similarly to Duchowski \cite{Duchowski2020}, we showed that LHIPA can be used as an indicator of difficulty or cognitive load by AOI. Additionally, we demonstrated that the visual attention reflected by the traditional metrics may not reflect the cognitive load. Considering the nose area, for example, it had the highest fixation time value but one of the lowest pupillary activity values. The robustness of the LHIPA response demonstrates its potential for further discussions about which areas in a visual stimulus are cognitively relevant to an observer for decision-making.

\subsubsection{Acknowledgements} This study was financed in part by FEI, the Brazilian Federal Agency CAPES - Finance Code 001, and the Sao Paulo Research Foundation FAPESP (2018/13076-9).


\end{document}